\documentclass[prd,12pt,showpacs,aps,nofootinbib]{revtex4-1}
\usepackage{amssymb}
\usepackage{amsmath,amssymb,graphicx}
\usepackage{graphicx}
\usepackage{dcolumn}
\usepackage{bm}
\def\be{\begin{equation}}
\def\ee{\end{equation}}
\def\bea{\begin{eqnarray}}
\def\eea{\end{eqnarray}}

\usepackage{axodraw4j}
\usepackage{axodraw}
\usepackage{pstricks}
\usepackage{graphicx}
\usepackage{amsbsy}
\usepackage{bm}
\usepackage{color}
\usepackage{fancyhdr}
\usepackage{epsfig}
\usepackage{hyperref}

\begin{document}

\bigskip

\vspace{2cm}
\title{Bounding resonant Majorana neutrinos from four-body $B$ and $D$ decays }
\vskip 6ex
\author{G. L\'opez Castro}
\email{glopez@fis.cinvestav.mx}
\author{N. Quintero}
\email{nquintero@fis.cinvestav.mx}
\affiliation{Departamento de F\'isica, Centro de Investigaci\'on y de Estudios Avanzados, Apartado Postal 14-740, 07000 M\'exico D.F., M\'exico}

\bigskip
\begin{abstract}
  Searches of lepton-number violation in different processes are very useful to constrain the parameter space of Majorana neutrinos. Here we use available upper bounds on the branching fractions of $B^-\to D^0\pi^+\mu^-\mu^-$ and $D^0\to (\pi^-\pi^-/K^-\pi^-)\mu^-\mu^-$ decays to derive constraints on the mass and mixings of Majorana neutrinos by assuming they are produced resonantly in these four-body decays. While the excluded region obtained from $B^-$ decays are competitive with existing limits from three-body $D^-$ and $B^-$ decays, it is shown that experimental improvements on $D^0$ decays offer a good potential to provide similar results.    

\end{abstract}

\maketitle
\bigskip

\section{Introduction}
If neutrinos turn out to be Majorana particles, their effects should manifest in lepton number violating (LNV) processes where the total lepton number changes in two units ($\Delta L=2$) \cite{Pontecorvo:1957qd,Valle:1982}. Conversely, the observation of decay/production phenomena with $\Delta L=2$ would be very helpful in elucidating the mechanism of neutrino mass generation \cite{cheng-li}.  Given its relevance, it is very important to study all possible channels that may be sensitive to the effects of $\Delta L=2$ interactions and explore the constrains that they provide on the parameter space of specific models.

  In the case that LNV are induced by the exchange of Majorana neutrinos, their masses and mixing angles can be constrained from the experimental upper limits of $\Delta L=2$ observables. It is very well known that the most sensitive channels to very light Majorana neutrinos are neutrinoless double beta decays of some nuclei \cite{doublebeta}. On the other hand, $\Delta L=2$ decays of pseudoscalar mesons \cite{Atre:2009,Helo:2011,Kovalenko2005,Cvetic:2010,Zhang:2011,Baoa:2012} and tau leptons \cite{Atre:2009,Helo:2011,LNVtau}  have proven to be useful to constrain sterile neutrinos with masses in the MeV to a few GeV range, which can be produced on their mass-shell \cite{Atre:2009} in these reactions. Majorana particles of this kind are known as {\it resonant} neutrinos \cite{Atre:2009}. In addition to these widely studied three-body decays \cite{Atre:2009,Helo:2011,Kovalenko2005,Cvetic:2010,Zhang:2011,Baoa:2012,LNVtau,PDG,babar,LHCb,belle2011,belle2010}, in previous papers we have reported results on the analysis of four-body decays of neutral mesons \cite{Quintero:2011} and tau leptons \cite{Quintero:2012a} (see also \cite{Dib:2012}). It has been shown \cite{Quintero:2011,Quintero:2012a,Quintero:2012b} that searches of these new (yet unexplored) decay channels can provide constraints on the parameter space of Majorana neutrinos that are complementary to three-body decays.  

  In the present Brief Report we study the constraints that can be gotten from current experimental bounds on four-body decays of heavy mesons. Our study is motivated by searches reported recently by the LHCb collaboration, namely \cite{LHCb}
  \be \label{LHCb4body}
\mathcal{B}(B^{-} \to D^{0}\pi^{+}\mu^{-}\mu^{-}) < 1.5\times10^{-6}. 
\ee
This decay channel may receive contributions from the exchange of heavy Majorana neutrinos via two different Feynman diagrams. We prove that in this case the Cabibbo-allowed channel, similar to the dominant contribution that underlies neutral $B$ meson decays \cite{Quintero:2011}, is favored over the Cabibbo-suppressed one in most of the neutrino mass region. 

In addition, we also study the constrains provided by the four-body LNV decay channels of $D^0$ mesons. Althought experimental searches for these decays were reported long ago by the E791 collaboration \cite{E791}, no theoretical studies have been done so far. The reported upper limits \cite{E791}
currently are very mild,  however our study shows that the constrains on the parameter space of Majorana neutrinos can become competitive and complementary to the ones gotten from three-body decays with improved limits on the branching fractions.

\section{Four-body LNV $B^-$ decays}
\begin{figure}[!h]
\centering
\includegraphics[scale=0.6]{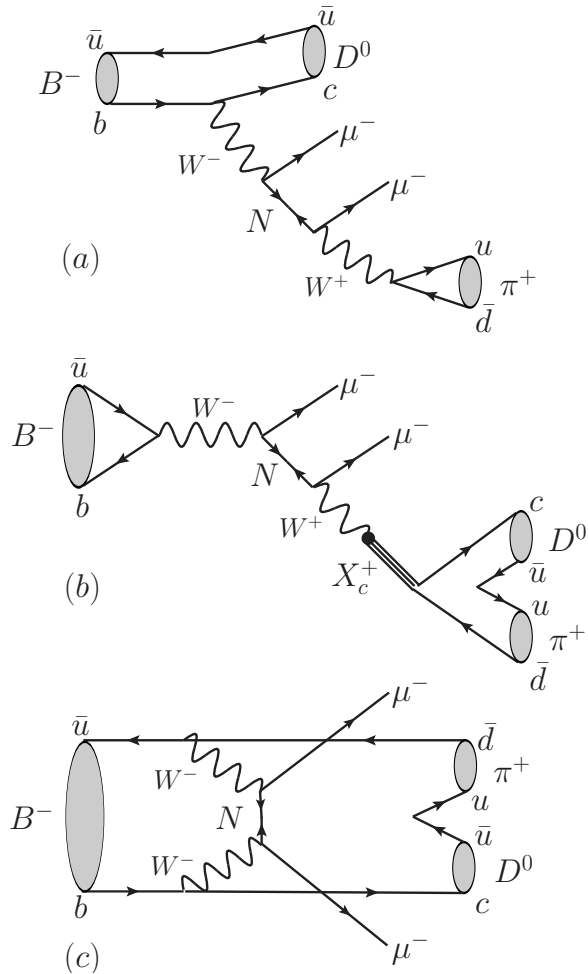}\\
\caption{\small Lowest order  $(a)$ spectator, ($b$) annihilation, and ($c$) $t$-channel diagrams, mediated by a heavy Majorana neutrino $N$ in $B^{-} \to D^{0}\pi^{+}\mu^{-}\mu^{-}$ decay.}
\label{Fig:Feynman}
\end{figure}

The Feynman diagrams that contribute to the LNV decay $B^{-}\to D^{0}\pi^{+} \mu^{-} \mu^{-}$ are shown in Fig. \ref{Fig:Feynman}. As in previous studies, we assume that only one heavy neutrino $N$, with a mass such that it can be produced resonantly in $B$ decays, dominates the decay amplitude. For such heavy neutrinos, the diagram of Figure 1c gives a negligible contribution \cite{Kovalenko2005}. 

  The amplitude in Figure 1a can be resonantly enhanced for neutrino masses in the range $m_{\pi}+m_{\mu} < m_N < m_B-m_D-m_{\mu}$. With obvious notation for particle momenta, it can be written as \cite{Quintero:2011,Quintero:2012b}
\be 
{\cal M}_{(a)}^{B^-}
= G_F^2 V_{cb}^{\text{CKM}}V_{ud}^{\text{CKM}} \langle D(p_D)|\bar{c}\gamma^{\alpha}b|B(p_B)\rangle [if_\pi p_\pi^{\beta}] L_{\alpha\beta} , \label{AmpliA}
\ee

\noindent where $|V_{cb}^{\text{CKM}}| = 40.9\times10^{-3}$ and $|V_{ud}^{\text{CKM}}|=0.97425$ \cite{PDG} are the Cabibbo-Kobayashi-Maskawa quark mixing matrix elements, $G_F$ is the Fermi constant and $f_{\pi}= 130.4$ MeV \cite{PDG} the pion decay constant. The information about the Majorana neutrino exchange is included in the leptonic tensor $L_{\alpha\beta}$ (properly antisymmetrized under exchange of lepton momenta $p_{1,2}$) \cite{Atre:2009}
\begin{eqnarray}\label{LeptI}
L_{\alpha\beta} = && |V_{\mu N}|^2 m_{N} \ \bar{u}(p_1) \Bigg[\frac{\gamma_{\alpha}\gamma_{\beta}}{(Q-p_1)^{2} - m_{N}^{2} + i \Gamma_{N}m_{N}} + (\alpha\leftrightarrow \beta, p_1 \leftrightarrow p_2) \Bigg]P_{R}u^{c}(p_2),
\end{eqnarray} 

\noindent with $P_R = (1+\gamma_{5})/2$, $Q=p_1+p_2+p_\pi$ the momentum transfer, $m_N$ and $\Gamma_N$ denote the mass and decay width of the heavy neutrino, respectively ($\Gamma_N$ can be calculated as shown in Ref. \cite{Atre:2009}). The constant $V_{\mu N}$ denotes the mixing of the muon with the heavy neutrino $N$ in the charged current interaction.

The hadronic matrix element in Eq. (\ref{AmpliA}) is given by
\begin{eqnarray}\label{BDtran}
\langle D(p_D)|\bar{c}\gamma_{\alpha}b| B(p_B)\rangle = && \Big((p_B + p_D)_\alpha -\frac{\Delta}{t} \ Q_{\alpha}\Big) F_1^{B\to D}(t) + \frac{\Delta}{t} Q_{\alpha}  F_{0}^{B\to D}(t),
\end{eqnarray}

\noindent where $\Delta\equiv (m^2_{B}-m^2_{D})$, $F_{0,1}^{B \to D}$  are the scalar and vector form factors for the $B \to D$ transition evaluated at $t=Q^{2}$. We will use the predictions for these form factors that are obtained from Lattice QCD calculations \cite{Kronfeld}. 

The contribution from Figure 1b is similar to the one encountered in three-body decays  $B^{-} \to X_c^{+} \mu^{-}\mu^{-}$ with the subsequent decay of the charmed resonance $X_c^+ \to D^0\pi^+$. The corresponding amplitude becomes resonantly enhanced for neutrino mass values in the range $m_D+m_{\pi}+m_{\mu} < m_N < m_B-m_{\mu}$.  The associated decay amplitude is given by
\be \label{AmpliB}
{\cal M}_{(b)}^{B^-}
= G_F^2 V_{ub}^{\text{CKM}}V_{cd}^{\text{CKM}} [i f_{B} p_B^{\alpha}] \ \langle D(p_D)\pi(p_\pi) |\bar{d}\gamma^{\beta}c|0\rangle \widetilde{L}_{\alpha\beta},
\ee

\noindent where $|V_{ub}^{\text{CKM}}| = 4.15 \times10^{-3}$, $|V_{cd}^{\text{CKM}}|=0.230$ \cite{PDG}, $f_B = 194$ MeV \cite{PDG} and the leptonic tensor 
\begin{equation}\label{LeptII}
\widetilde{L}_{\alpha\beta} = |V_{\mu N}|^2 m_{N} \ \bar{u}(p_1) \Bigg[\frac{\gamma_{\alpha}\gamma_{\beta}}{(p_B-p_1)^{2} - m_{N}^{2} + i \Gamma_{N}m_{N}} + (\alpha\leftrightarrow \beta, p_1 \leftrightarrow p_2) \Bigg]P_{R}u^{c}(p_2).
\end{equation}

\noindent The hadronic matrix element  is parametrized as follows
\begin{eqnarray}
\langle D(p_D)\pi(p_\pi) |\bar{d}\gamma^{\alpha}c|0\rangle  &=& \Big( (p_D - p_\pi)^{\alpha}  +\frac{\Delta'}{k^2}k^{\alpha}\Big) F_{1}^{D\to \pi}(k^2)+\frac{\Delta'}{k^2}k^{\alpha}F_0^{D\to \pi}(k^2), 
\end{eqnarray} 

\noindent where $k=p_D + p_\pi$ and $\Delta'=m_D^2-m_{\pi}^2$. The form factors will be modeled as $F_{1,0}^{D\to \pi}(k^2)= F_{1,0}^{D\to \pi}(0)\cdot BW_{X_c}(k^2)$, with $F_{1,0}^{D\to \pi}(0) = 0.67$ \cite{CLEO} its value at zero momentum transfer. We will use a simplified model where the Breit-Wigner (BW) function introduced above is dominated by a single resonance:
\be
BW_{X_c}(k^{2}) = m_{X_c}^{2}/ \left[m_{X_c}^{2} - k^{2} - i m_{X_c}\Gamma_{X_c}\right],
\ee

\noindent where the mass and width of the charmed resonances $X_c$ correspond to the $D^{*+}(2010)$ and $D_{0}^{*+}(2400)$, respectively, for the spin-1 and spin-0 form factors.  

Althought the annihilation amplitude in Figure 1b is Cabibbo-suppressed with respect to the spectator amplitude from Fig. 1a, a strong enhancement of the former due to the exchange of the narrow $D^*(2010)$ resonance is possible. Note that the sensitivity to the mass of Majorana neutrinos explored by diagrams 1a and 1b are complementary and overlap only over a small window $2.1\ {\rm GeV} \leq m_N \leq 3.3\ {\rm GeV}$.

\begin{figure}[!h]
\centering
\includegraphics[scale=0.6]{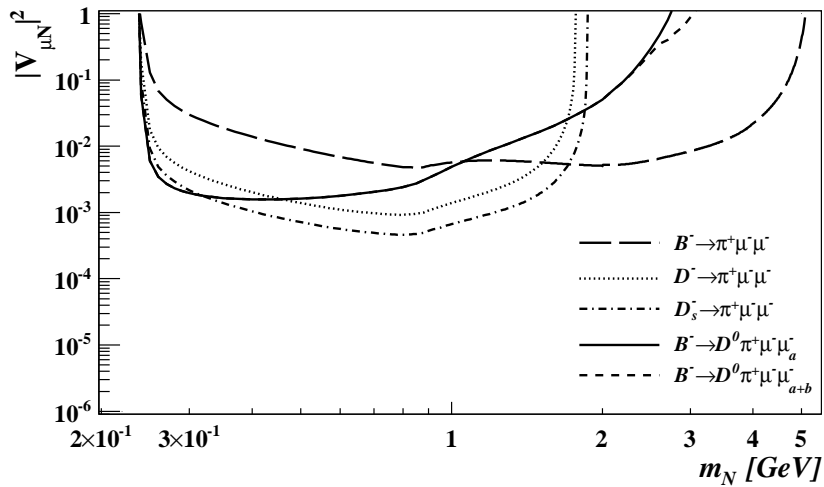}\\
\caption{\small Exclusion regions on heavy mixing $|V_{\mu N}|^{2}$ as a function of the Majorana neutrino mass $m_N$, from LNV dimuon modes of heavy mesons. The solid [short-dashed] lines denote the constraints obtained from $B^- \to D^0\pi^+\mu^-\mu^-$ by including diagrams $(a)$ [$(a)+(b)$] from Figure 1. }
\label{VmuNB}
\end{figure}

  The phase space of the four-body decay is determined by the limits on the five independent kinematical variables \cite{Asatrian:2012,FloresTlalpa:2005fz}. We perform the numerical integrations using the \texttt{VEGAS} code \cite{Lepage:1977sw} and implementing the single-diagram enhanced channel integration method \cite{Maltoni}.
 Figure \ref{VmuNB} shows the excluded region (above the solid and short-dashed curves) for  $|V_{\mu N}|^{2}$ as a function of $m_N$ that is obtained from the upper limit reported by the LHCb Collaboration, Eq. (\ref{LHCb4body}). As it was pointed out above, the dominant effects of Majorana neutrinos come the diagram in Figure 1a while the effects from diagram 1b are visible only at higher values of $m_N$. 
For comparison, we also display the updated exclusion plots obtained from searches of three-body $(D^{-}, D_s^{-}, B^{-}) \to \pi^{+}\mu^{-}\mu^{-}$ decays. For neutrino mass values below 1 GeV, the four-body decay under consideration is able to exclude a larger region of the $|V_{\mu N}|^{2}$ than the Cabibbo-suppressed channel $B^- \to \pi^+\mu^-\mu^-$ (long-dashed plot in Figure 2) despite the fact that the upper limit on the later is of order $10^{-8}$ \cite{LHCb}. Therefore, future and improved upper limits on the ${\cal B}(B^- \to D^0\pi^+\mu^-\mu^-)$ will be very useful to derive better constrains on the $\mu N$ mixing angle.

An analogous process to the one studied before is the  $D^- \to \pi^0\pi^+\mu^{-}\mu^{-}$ decay. If this $\Delta L=2$ decay is due to the exchange of Majorana neutrinos, we should have similar contributions to the amplitude as the ones shown in Figure 1.  The corresponding spectator and annihilation contributions of Figs 1a and 1b are both proportional to the product $V_{cd}^{\text{CKM}}V_{ud}^{\text{CKM}}$, therefore we can expect that the $\rho^+(770)$ resonance intermediate contribution (Fig. 1b) becomes important in this case. Moreover, the sensitivity region of neutrino masses where spectator and annihilation amplitudes are resonantly enhanced are very similar. However, since this four-body decay has similar Cabibbo-suppressed couplings as $D^- \to \pi^+ \mu^-\mu^-$ we expect that the excluded region on Majorana neutrinos from experimental upper limits on both decays would be very similar.

\section{Four-body LNV $D^0$ decays}
The four-body $D^0 \to h^-h'^-l^+l'^+$ decays, where $h,h'=\pi,\ K$ and $l,l'=e, \mu$, were studied by the E791 collaboration \cite{E791} more than a decade ago and 90\% CL  upper limits of order $10^{-4}\sim 10^{-5}$ for their branching ratios were obtained. In this section we focus on the $D^{0}\to (K^-\pi^-,\pi^-\pi^-) \mu^{+} \mu^+$ channels in order to illustrate the potential these decays offer to constrain the parameters of Majorana neutrinos. The upper limits for these decays obtained in Ref. \cite{E791} are:
\begin{eqnarray}
\mathcal{B}(D^0 \to \pi^-\pi^-\mu^+\mu^{+}) &<& 2.9\times 10^{-5}, \nonumber \\
\mathcal{B}(D^0 \to K^-\pi^-\mu^+\mu^{+}) &<& 3.9\times 10^{-4}. \label{d0limits}
\end{eqnarray}

Since weak currents carry one unit of electric charge, the Feynman diagram analogous to Figure 1b does not contribute in this case. The dominant contribution to the decay amplitude is given  by the spectator diagram similar to Figure 1a and its amplitude becomes:
\be 
{\cal M}_{(a)}^{D^0}
= G_F^2 V_{cq}^{\text{CKM}}V_{ud}^{\text{CKM}} \langle h(p_{h})|\bar{q}\gamma^{\alpha}c|D(p_D)\rangle [if_\pi p_\pi^{\beta}] L_{\alpha\beta},
\ee

\noindent with $q=d(s)$ for $h = \pi (K)$. The leptonic tensor current $L_{\alpha\beta}$ is given by Eq. (\ref{LeptI}) and the hadronic current $\langle h(p_{h})|\bar{q}\gamma^{\alpha}c|D(p_D)\rangle$ can be parametrized as in Eq. (\ref{BDtran}). For the purposes of our numerical evaluation we take the form factors $F_{0,1}^{D\to h}$ from Ref. \cite{CLEO,Lu:2011}. In the case of identical mesons in the final state the decay amplitude must, in addition, be symmetrized under exchange of their momenta. 

\begin{figure}[!h]
\centering
\includegraphics[scale=0.6]{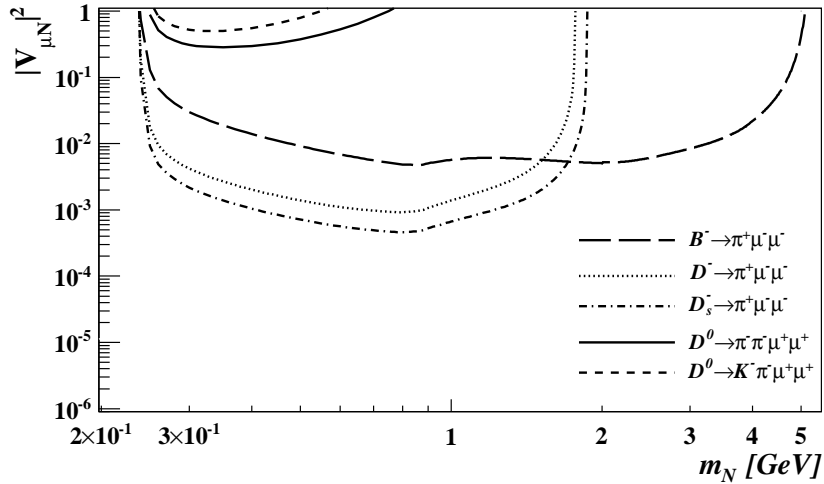}\\
\caption{\small Exclusion curves for $|V_{\mu N}|^2$ as a function of $m_N$ from four-body LNV $D^0$ decays (solid and short-dashed lines). Constraints from other heavy meson decays are shown for comparison.}\label{VmuN_D}
\end{figure}

 By using the upper limits given in (\ref{d0limits}), in Fig. \ref{VmuN_D} we plot the exclusion regions provided by $D^0 \to \pi^-\pi^-\mu^+\mu^{+}$ (solid line) and $D^0 \to K^-\pi^-\mu^+\mu^{+}$ (short-dashed line) decays. Currently, these $D^0$ decays provide only loose constrains on Majorana neutrinos given the poor upper limits available on branching ratios.  Note however that improvements by two or three orders of magnitude on the branching ratios would yield competetive constraints on the mixing angle $|V_{\mu N}|^2$ compared to other LNV $D^{\pm}$ meson decays, as it can be appreciated in Figure \ref{VmuN_D}. This goal is certainly  at the reach of current (LHCb, BESIII) and future (Belle II) $D$ meson factories. 

\bigskip

Summarizing, in this Report we derive the first constraints on parameters of Majorana neutrinos that can be obtained from the LNV decays  $B^- \to D^0 \pi^+ \mu^-\mu^-$ and  $D^0 \to (K^-\pi^-/\pi^-\pi^-)\mu^+\mu^+$.  We use the upper limits on the corresponding branching ratios that have been reported, respectively,  by the LHCb \cite{LHCb} and the E791 \cite{E791} Collaborations.  As in previous works, we assume that only one Majorana neutrino $N$, with a mass of hundreds of MeV to a few GeV, contribute resonantly to the decay amplitude (LNV heavy meson decays are insensitive to either, very light or very heavy neutrinos \cite{Ali:2006}). 
  The excluded region in the $|V_{\mu N}|^2$ vs. $m_N$ plane obtained from the upper limit on ${\cal B}(B^- \to D^0 \pi^+ \mu^-\mu^-)$ reported by LHCb \cite{LHCb} is similar to the one obtained from existing bounds on LNV three body decays of charged $D$ and $B$ mesons. This is possible because the four-body decay is a Cabibbo-allowed decay, compared to the three-body decay which proceeds via a Cabibbo-suppressed mechanism. 

  We have also studied the four-body LNV decays of the neutral charmed meson. We use the two most restrictive upper bounds obtained by the E791 collaboration to derive bounds on the same parameters of Majorana neutrinos. These bounds turn out to be very mild at present, but their improvements at current and planned $D$ meson factories would make them competitive with constrains derived from three-body decays of charged mesons. 

 Finally, let us comment that stronger constraints on the $|V_{\mu N}|$ mixing for the same range of sterile neutrino masses can be obtained from searches of peaks in the muon spectrum of leptonic $K^{\pm}$ decays \cite{Kusenko:2004qc} or from searches of specific visible channels of heavy neutrino decays produced in beam dump or neutrino scattering experiments \cite{other}. The non-observation of these signals allows to put constraints on $|V_{\mu N}|^2$ up to $10^{-6}$ or $10^{-8}$  for masses of the  sterile neutrino ranging from 0.2 to 2 GeV \cite{Atre:2009,Kusenko:2004qc,other}. In comparison, the current constraints obtained from direct searches of ($\Delta L=2$) LNV decays, including the ones considered in this Report, are certainly less restrictive except for the $K^+ \to \pi^- \mu^+\mu^+$ decays which can reach $|V_{\mu N}|^2\sim 10^{-8}$ but only for a very narrow range ($0.25\ {\rm MeV} \leq m_N \leq 0.38$ MeV) of Majorana neutrino masses. Similarly, LNV decays of $B$ mesons may eventually provide better better constraints on $|V_{\mu N}|^2$ for $2 \leq m_N \leq 5$ GeV than those obtained from heavy neutrino decays of $Z^0$ bosons \cite{LEP} ($|V_{\mu N}|^2 \leq 10^{-4}$).  

\medskip

The authors would like to thank Conacyt (M\'exico) for financial support.



\begin{thebibliography}{99}
\bibitem{Pontecorvo:1957qd}
  B.~Pontecorvo,
  Sov.\ Phys.\ JETP {\bf 7}, 172 (1958); 
  Sov.\ Phys.\ JETP {\bf 26}, 984 (1968); 
  V.~N.~Gribov and B.~Pontecorvo,
  Phys.\ Lett.\ B {\bf 28}, 493 (1969).

\bibitem{Valle:1982}
J. Schechter and J. W. F. Valle, Phys. Rev. D \textbf{25}, 2951 (1982).

\bibitem{cheng-li}
T. P. Cheng and L. F. Li, Phys. Rev. D{\bf 22}, 2860 (1980); W. Konetschny and W. Kummer, Phys. Lett. {\bf B70}, 433 (1977); G.~Lazarides, Q.~Shafi and C.~Wetterich,  Nucl.\ Phys.\ B {\bf 181}, 287 (1981);  J.~Schechter and J.~W.~F.~Valle,
  Phys.\ Rev.\ D {\bf 22}, 2227 (1980);  R.~N.~Mohapatra and G.~Senjanovic,
  Phys.\ Rev.\ D {\bf 23}, 165 (1981).

\bibitem{doublebeta}
  M.~Doi, T.~Kotani, H.~Nishiura, K.~Okuda, E.~Takasugi,
  Phys.\ Lett.\  {\bf B102}, 323 (1981);  S.~R.~Elliott and P.~Vogel,
  Ann.\ Rev.\ Nucl.\ Part.\ Sci.\  {\bf 52}, 115 (2002); 
  H.~V.~Klapdor-Kleingrothaus,  {\it et al.},
  Eur.\ Phys.\ J.\  {\bf A12}, 147 (2001). 
  For a recent review see: W.~Rodejohann,
  Int.\ J.\ Mod.\ Phys.\ E {\bf 20}, 1833 (2011).

\bibitem{Atre:2009}
A. Atre, T. Han, S. Pascoli, and B. Zhang, JHEP \textbf{05}, 030 (2009).  



\bibitem{Helo:2011}
J. C. Helo, S. Kovalenko, and I. Schmidt, Nucl. Phys. \textbf{B853}, 80 (2011).

\bibitem{Kovalenko2005}
M. A. Ivanov and S. G. Kovalenko, Phys. Rev. D \textbf{71}, 053004 (2005).

\bibitem{Cvetic:2010}
  G.~Cvetic, C. Dib, S.~K.~Kang, and C.~S.~Kim,
  Phys. Rev. D {\bf 82}, 053010 (2010); C. Dib, V. Gribanov, S. Kovalenko, and I. Schmidt, Phys. Lett. B \textbf{493}, 82 (2000).

\bibitem{Zhang:2011}
  J.~M.~Zhang and G.~L.~Wang,
  Eur.\ Phys.\ J.\  C {\bf 71}, 1715 (2011).  
    
\bibitem{Baoa:2012}
S.-S. Bao, H.-L. Li, Z.-G. Si, and Y.-B. Yang, Commun.\  Theor.\  Phys.\  {\bf 59}, 472 (2013).

\bibitem{LNVtau}
A. Ilakovac, Phys. Rev. D \textbf{54}, 5653
(1996); V. Gribanov, S. Kovalenko, and I. Schmidt, Nucl.
Phys. \textbf{B607}, 355 (2001).


\bibitem{PDG} 
J. Beringer \textit{et al.} [Particle Data Group], Phys. Rev. D \textbf{86}, 010001 (2012).

\bibitem{babar}
J. P. Lees {\it et al.} [BABAR Collaboration], Phys. Rev. D \textbf{84}, 072006 (2011); \textbf{85}, 071103(R) (2012). 

\bibitem{LHCb} 
R. Aaij \textit{et al.} [LHCb Collaboration], Phys. Rev. Lett. \textbf{108}, 101601 (2012) 
; Phys. Rev. D \textbf{85}, 112004 (2012). 

\bibitem{belle2011}
O. Seon {\it et al.} [Belle Collaboration], Phys. Rev. D {\bf 84}, 071106(R) (2011).

\bibitem{belle2010}
Y. Miyazaki {\it et al.} [Belle Collaboration], Phys. Lett. B \textbf{682}, 355 (2010); arXiv:1206.5595 [hep-ex]. 
  


\bibitem{Quintero:2011}
D. Delepine, G. L\'opez Castro, and N. Quintero,  Phys. Rev. D \textbf{84}, 096011 (2011) [Erratum-\textit{ibid} D \textbf{86}, 079905 (2012)].


\bibitem{Quintero:2012a}
G. L\'opez Castro and N. Quintero,  Phys. Rev. D \textbf{85}, 076006 (2012) [Erratum-\textit{ibid} D \textbf{86}, 079904 (2012)].


\bibitem{Dib:2012}
C. Dib, J. C. Helo, M. Hirsch, S. Kovalenko, and I. Schmidt, Phys. Rev. D \textbf{85}, 011301(R) (2012).

\bibitem{Quintero:2012b}
G. L\'opez Castro and N. Quintero, arXiv:1212.0037 [hep-ph].


\bibitem{E791}
E. M. Aitala {\it et al.} [E791 Collaboration], Phys. Rev. Lett. \textbf{86}, 3969 (2001).



\bibitem{Kronfeld}
Jon A. Bailey {\it et al.} [Fermilab Lattice and MILC], Phys. Rev. D \textbf{85}, 114502 (2012) [Erratum-\textit{ibid} \textbf{86}, 039904 (2012)]. 


\bibitem{CLEO}
D. Besson \textit{et al.} [CLEO Collaboration], Phys. Rev. D \textbf{80}, 032005 (2009).



\bibitem{FloresTlalpa:2005fz} 
  A.~Flores-Tlalpa, G.~Lopez Castro and G. Toledo Sanchez,
  Phys.\ Rev.\ D {\bf 72}, 113003 (2005).


\bibitem{Asatrian:2012}
H. M. Asatrian, A. Hovhannisyan, and A. Yeghiazaryan, Phys. Rev. D \textbf{86}, 114023 (2012).

\bibitem{Lepage:1977sw} 
  G.~P.~Lepage,
  J.\ Comput.\ Phys.\  {\bf 27}, 192 (1978).

\bibitem{Maltoni}
F. Maltoni and T. Stelzer, JHEP \textbf{02}, 027 (2003).

\bibitem{Lu:2011}
F.-S. Yu, X.-X. Wang, and C.-D. L\"u, Phys. Rev. D \textbf{84}, 074019 (2011).


\bibitem{Ali:2006}  
A.~Ali, A.~V.~Borisov and M.~V.~Sidorova, 
Phys.\ Atom.\ Nucl.\  {\bf 69}, 475 (2006)
[Yad.\ Fiz.\  {\bf 69}, 497 (2006)]; 
A.~Ali, A.~V.~Borisov and N.~B.~Zamorin, 
Eur. Phys. J. C {\bf 21}, 123 (2001); 
A.~Atre, V.~Barger and T.~Han, 
Phys. Rev. D {\bf 71}, 113014 (2005). 

\bibitem{Kusenko:2004qc} 
  A.~Kusenko, S.~Pascoli, D.~Semikoz and ,
  JHEP {\bf 0511}, 028 (2005).

\bibitem{other}
G.~Bernardi, {\it et al.},
  Phys.\ Lett.\ B {\bf 203}, 332 (1988); P.~Vilain {\it et al.}  [CHARM II Collaboration],
  Phys.\ Lett.\ B {\bf 343}, 453 (1995)
  [Phys.\ Lett.\ B {\bf 351}, 387 (1995)]; A.~Vaitaitis {\it et al.}  [NuTeV and E815 Collaborations],
  Phys.\ Rev.\ Lett.\  {\bf 83}, 4943 (1999); E.~Gallas {\it et al.}  [FMMF Collaboration],
  Phys.\ Rev.\ D {\bf 52}, 6 (1995); J.~Badier {\it et al.}  [NA3 Collaboration],
  Z.\ Phys.\ C {\bf 31}, 21 (1986).
\bibitem{LEP}
O. Adriani {\it et al} [L3 Collaboration], Phys. Lett. {\bf B95}, 371 (1992); P. Abreu {\it et al}, Z. Phys. {\bf C74}, 57 (1997).



  
\end{thebibliography}
\end{document}